\documentclass[12pt]{article}
\pdfoutput=1

\usepackage{geometry,amsmath,amsfonts}
\usepackage{slashed}
\usepackage{epsfig}
\usepackage{latexsym}
\usepackage{graphicx}
\usepackage{amssymb}
\usepackage[caption=false]{subfig}
\usepackage{color}
\usepackage{multirow}
\usepackage{color}
\usepackage{rotating}
\usepackage{ifthen}
\usepackage{epsfig}
\usepackage{lineno}
\usepackage{cite}

\newcommand{\beq}{\begin{equation}}
\newcommand{\eeq}{\end{equation}}
\newcommand{\bea}{\begin{eqnarray}}
\newcommand{\eea}{\end{eqnarray}}

\newcommand{\gsim}{\lower.7ex\hbox{$\;\stackrel{\textstyle>}{\sim}\;$}}
\newcommand{\lsim}{\lower.7ex\hbox{$\;\stackrel{\textstyle<}{\sim}\;$}}

\begin{document}

\vspace*{10mm}

\begin{center}

{\large\bf  Models with two Higgs doublets and a light pseudoscalar:}

\vspace*{3mm}

\mbox{\large\bf a portal to dark matter and the possible $\mathbf{(g-2)_\mu}$ excess} 

\vspace*{12mm}

{\sc Giorgio Arcadi$^1$},  {\sc Abdelhak~Djouadi$^{2,3}$} and {\sc Farinaldo Queiroz$^{4,5,6}$} 

\vspace*{12mm}

{\small

$^1$ Dipartimento di Scienze Matematiche e Informatiche, Scienze Fisiche e Scienze della Terra, \\ Universita degli Studi di Messina, Via Ferdinando Stagno d'Alcontres 31, I-98166 Messina, Italy. 
\\ \vspace{0.2cm}

$^2$ CAFPE and Departamento de Fisica Te\'orica y del Cosmos,\\ Universidad de Granada, E--18071 Granada, Spain.\\ \vspace{0.2cm}

$^3$ NICPB, R{\"a}vala pst. 10, 10143 Tallinn, Estonia.\\ \vspace{0.2cm}

$^4$ Departamento de F\'isica, Universidade Federal do Rio Grande do Norte, 59078-970, Natal, RN, Brasil.\\  \vspace{0.2cm}

$^5$ International Institute of Physics, Universidade Federal do Rio Grande do Norte, Campus Universitário, Lagoa Nova, Natal-RN 59078-970, Brazil. \\ \vspace{0.2cm}

$^6$  Millennium Institute for Subatomic Physics at High-Energy Frontier (SAPHIR),
Fernandez Concha 700, Santiago, Chile. \\ \vspace{0.2cm}.

}

\end{center}

\vspace*{12mm}

\begin{abstract} 

In the context of a two-Higgs doublet model, supplemented by an additional light pseudoscalar Higgs boson and a stable isosinglet fermion, we consider the possibility of addressing simultaneously the discrepancy from the standard expectation of the anomalous magnetic moment of the muon recently measured at Fermilab and the 
longstanding problem of the dark matter in the universe which can be accounted for by a thermal weakly interacting massive particle. We show that it is indeed possible, for a range of masses and couplings of the new light pseudoscalar and the fermionic states, to explain at the same time the two features while  satisfying all other constraints from astroparticle physics and collider searches, including the  constraints from flavor physics.

\end{abstract}

\newpage

\subsection*{1. Introduction}
 
 A new measurement of the anomalous magnetic moment of the muon, $a_\mu= \frac12 (g-2)_\mu$, has been recently released by the Muon $g-2$ collaboration at Fermilab~\cite{Abi:2021gix} which, when combined with a previous measurement performed at Brookhaven ~\cite{Bennett:2006fi}, gives the  value~\cite{Abi:2021gix}
\beq \label{eq:4sigma}
    a_\mu^{\rm EXP}= (116 592 061 \pm 41)\times10^{-11},
\eeq
which implies a $4.2 \sigma$ deviation from the consensus on the Standard Model (SM) contribution~\cite{Aoyama:2020ynm}
\beq
\Delta a_\mu= a_\mu^{\rm EXP} - a_\mu^{\rm SM} = (251 \pm 59)\times10^{-11} \, .
\eeq

While not yet exceeding the 5$\sigma$ target which is needed to claim observation,  it is very tempting to attribute this discrepancy to a new phenomenon beyond the ones predicted in the SM \cite{Lindner:2016bgg}, rather to still unknown theoretical or experimental uncertainties. In such  a case, the new measurement would probably be the first sign of the so awaited new physics. An explanation of the fact that these possible effects behind this observation have not been observed in direct searches conducted in the high-energy frontier at the CERN Large Hadron Collider (LHC) 
\cite{Yuan:2020fyf,Bailey:2020ecs,Queitsch-Maitland:2020oxk}, would be that they are rather due to the presence of light new species which can significantly contribute to the $(g-2)_\mu$ observable, but are difficult to detect at the LHC as they yield events with small transverse momenta.  The new light degrees of freedom could also enter B-meson physics observables (in which some anomalies have also been observed) and in, particular, contribute to  the semi-leptonic $b \to s\mu^+\mu^-$ decay rate, which  happens to also be related to muons  and slightly deviates from the SM expectation \cite{LHCb:2021awg}. 

All these anomalies, if they are indeed present, need  to be related to the other puzzle that we have in particle physics, namely the presence of dark matter (DM) in the universe \cite{Planck:2018vyg}. This DM could appear in the form of a colorless and electrically neutral, weakly interacting massive particle (WIMP)  which is stable at cosmological scales \cite{Bertone:2004pz,Arcadi:2017kky}. Several attempts have been made in this direction, see e.g. Refs.~\cite{Lu:2021vcp,Chowdhury:2021tnm,Arcadi:2021cwg,Arcadi:2021yyr,Hapitas:2021ilr,Acuna:2021rbg,Arcadi:2021glq} for a few  examples. In Ref.~\cite{Arcadi:2021cwg} for instance, a systematic classification of minimal models according to the quantum numbers of their field content \cite{Calibbi:2018rzv} has been made and two specific examples of scenarios resolving the $(g-2)_\mu$ anomaly and with different DM candidates have been proposed: a mixed ${\rm SU(2)_L}$ singlet-doublet lepton  and a real scalar field. In Refs.~\cite{Arcadi:2021yyr,Hapitas:2021ilr}, a two Higgs doublet model (2HDM) augmented by an abelian gauge symmetry and a vector-like  fermion family, that contribute to  $(g-2)_\mu$ has also been discussed. 

In this note, we propose another solution to the $(g-2)_\mu$ possible  discrepancy which also fulfills the requirements for a WIMP dark matter. It is based on a Higgs sector which is extended to contain two-Higgs doublets and a light pseudoscalar field $a$ which can serve as a portal to a DM sector, which minimally consists of an SU(2) isosinglet fermion. This 2HD+$a$ model\footnote{In fact, this model bears many similarities with a well known benchmark scenario proposed for the next-to-minimal supersymmetric extension of the SM (NMSSM) \cite{Djouadi:2008uw}; see Ref.~\cite{Arcadi:2019lka} for a comparison.} has recently gained  a wide interest as it easily copes with constraints from collider and astroparticle physics \cite{Ipek:2014gua,Goncalves:2016iyg,Bauer:2017ota,Tunney:2017yfp,Abe:2018bpo}. Indeed, one can obtain the correct relic density for the DM through its efficient annihilation into SM (as well as the $a$) particles via the $s$-channel exchange of the $a$ state and, at the same time,  evade the stringent XENON1T direct limits in the spin-independent scattering of the DM over nucleons \cite{XENON:2018voc}, as the DM would not couple to the CP-even Higgs bosons. On the other hand, light pseudoscalar Higgs particles that do not couple strongly to the observed SM-like Higgs boson, can easily evade the LHC bounds from direct Higgs searches \cite{Bailey:2020ecs}.

Hence, the pseudoscalar particle present in the model can be rather light and have couplings to isospin-down fermions that are enhanced; it can be thus exchanged between muons and gives a contribution to the $(g-2)_\mu$ \cite{Dedes:2001nx,Chang:2000ii,Larios:2001ma,Ilisie:2015tra,Ferreira:2021gke,Jueid:2021avn,Eung:2021bef}. Whether this contribution is large enough as to explain the excess observed by the Fermilab experiment, while complying with the set of astrophysical and collider constraints previously mentioned, is the purpose of the present note.
 We will show that, indeed, there is a range of the  masses and couplings of the $a$ boson and the DM fermion that are not excluded by searches at the LHC and elsewhere and by direct and indirect detection experiments, which lead to the correct DM relic abundance and explains the $(g-2)_\mu$ deviation. In addition, the $a$ state would also contribute to the $b\! \to\! s l^+\l^-$ process which can be observed in B-meson decays; the decay rate would be particularly enhanced in the case of a light $a$ boson which is emitted on mass shell. 

In the next section, we briefly introduce the 2HD+$a$ model and summarize the theoretical constraints on it. In section 3, we summarize the various experimental constraints from LHC and other collider searches, DM experiments and the contributions to the $(g-2)_\mu$. In section 4, we perform a numerical analysis of the model and delineate the region of parameter space that could explain all observed phenomena. A short conclusion is given in the last section.  

%%%%%%%%%%%%%%%%%%%%%%%%%%%%%%%%%%%%%%%%%%%%%%%%%%%%%%%%%%%%%%%%%%%
\subsection*{2. The 2HD+a model}

The scenario of a two-Higgs doublet model plus a light pseudoscalar state offers the possibility of inducing in a gauge invariant manner, an interaction between a singlet pseudoscalar $a$ boson and the SM fermions. One obtains a coupling of the form $a \bar f \gamma_5 f$, via the mixing of $a$ with the pseudoscalar $A$ state of the 2HDM \cite{Ipek:2014gua,Goncalves:2016iyg,Bauer:2017ota,Tunney:2017yfp,Abe:2018bpo}. The most general scalar potential for such a model is given by~\cite{Abe:2018bpo}
\begin{equation} 
V = V(\Phi_1,\Phi_2) + \frac{1}{2} m_{a_0}^2 a_0^2+\frac{\lambda_a}{4}a_0^4
+\left(i \textcolor{red}{\kappa }  a_0 \Phi^{\dagger}_1\Phi_2+\mbox{h.c.}\right)+\left(\lambda_{1P}a_0^2 \Phi_1^{\dagger}\Phi_1+\lambda_{2P}a_0^2 \Phi_2^{\dagger}\Phi_2\right),
\end{equation} 
where $V(\Phi_1,\Phi_2)$ denotes the usual potential of the two Higgs doublet fields which can be found in  Refs.~\cite{Branco:2011iw,Djouadi:2005gj}. Once the electroweak symmetry is broken, the two doublet fields acquire non-zero expectation values $v_1$ and $v_2$ where, as usual, the ratio is denoted by $v_1/v_2=\tan\beta$ with $\sqrt{v_1^2+v_2^2}=v\simeq 246~{\rm GeV}$. The scalar sector of the theory will consist of two CP--even $h,H$ states, with $h$ conventionally identified with the 125 GeV boson observed at the LHC, two charged $H^{\pm}$ bosons and two CP--odd states. The latter are a mixture of the original singlet and 2HDM states  $a_0$ and $A_0$ obtained from the field rotation
\begin{equation}
\left(
\begin{array}{c} A \\ a \end{array} \right)= \begin{pmatrix}\! \cos\theta & \sin\theta \! \\ \!\! - \sin\theta & \cos\theta \! \end{pmatrix}\ \left( \begin{array}{c} A_0 \\ a_0 \end{array} \right) \, \quad 
{\rm with} \quad \tan2\theta=\frac{2 \kappa v}{M_{A}^2-M_{a}^2}\;.
\end{equation}

In the physical mass basis, the scalar sector of the theory is fully described by the following set of parameters: the physical masses of the five Higgs bosons, $M_h$, $M_H$, $M_{H^{\pm}}$, $M_a$, $M_A$, three parameters of the scalar potential, namely $\lambda_{1P}$, $\lambda_{2P}$ and $\lambda_3$ (contained in the 2HDM potential), and finally, the mixing angles entering the quantities $\sin\theta$, $\tan\beta$ and $\cos(\beta-\alpha)$ with $\alpha$ being the mixing angle among the 2HDM CP-even neutral bosons. It is possible to eliminate the last parameter by imposing the alignment limit, $\beta-\alpha=\pi/2$, which sets the values of the coupling of the lighter $h$ state to fermions and gauge bosons  to its corresponding SM values, as favored by the constraints on the 125 GeV Higgs boson properties measurements \cite{Yuan:2020fyf}. In addition, to cope with constraints from high-precision electroweak measurements performed at LEP and elsewhere \cite{deBlas:2021wap}, and in particular to forbid large contributions to the $\rho$ parameter, we will assume mass degeneracy for the heavier $H,A,H^{\pm}$ states, $M_H=M_{A}=M_{H^\pm}$ \cite{Haller:2018nnx}.

The couplings of the physical neutral Higgs bosons to the SM fermions play a crucial role in our context. They are described by the following Lagrangian
\begin{equation}
\mathcal{L}_{\rm Yuk}=\sum_f \frac{m_f}{v}\bigg[g_{hff} h \bar f f+g_{Hff}
H\bar f f- i g_{Aff} A  \bar f \gamma_5 f-i g_{aff} a \bar f \gamma_5 a \bigg] \, , \end{equation}
where, according to the adopted alignment limit $\alpha\!=\!\beta\! -\! \pi/2$, one should set the $h$ couplings to their SM values,  $g_{hff}\!=\!1$. For the other Higgs couplings, in order to avoid the appearance of flavor-changing neutral currents at tree-level, one assumes the following structure for them (the couplings of the charged Higgs bosons to isospin $\pm \frac12$ fermions follow that of the $H$ state) 
\begin{equation}
    g_{Hff}=\xi_f,\,\,\,\,g_{Aff}=\cos\theta \; \xi_f,\,\,\,\,\,\,g_{aff}=-\sin\theta \; \xi_f
\end{equation}
with the parameters $\xi_f$ having four sets of possible assignments, corresponding to four ``types" of 2HDM \cite{Branco:2011iw} and that are summarized in Table ~\ref{2hdm_type}.

\begin{table}[h!]
\renewcommand{\arraystretch}{1.2}
\begin{center}
\begin{tabular}{|c|c|c|c|c|}
\hline
~~~~~~ &  ~~Type I~~ & ~~Type II~ & Lepton-specific & Flipped \\
\hline\hline 
$g_{u}$ & $\frac{1}{\tan\beta}$ & $\frac{1}{\tan\beta}$ &$\frac{1}{\tan\beta}$ & $\frac{1}{\tan\beta}$\\
\hline
$g_{d}$ & $-\frac{1}{\tan\beta}$ & ${\tan\beta}$ & $-\frac{1}{\tan\beta}$ & ${\tan\beta}$ \\ \hline
$g_{l}$ & $-\frac{1}{\tan\beta}$ & ${\tan\beta}$ & ${\tan\beta}$ & $-\frac{1}{\tan\beta}$ \\ \hline
\end{tabular}
\caption{Summary of the possible values, in the alignment limit $\beta \!-\! \alpha \rightarrow \frac{\pi}{2}$, of the $\xi_f$ parameters describing the couplings of the extra Higgs bosons to the SM fermions.}
\label{2hdm_type}
\end{center}
\vspace*{-6mm}
\end{table}

Among these assignments, only the Type-II and the lepton-specific (also customarily dubbed Type-X) scenarios are of interest for our study, as they feature enhanced couplings of the additional Higgs bosons to the SM charged leptons for large values of $\tan\beta$.

The trilinear interactions between the Higgs states will also have an important impact. In the alignment limit, the pseudoscalar states couple only to the SM-like $h$ boson and an important interaction is the one among the $haa$ states described by the coupling 
\beq
\label{eq:haa}
  \lambda_{haa}\!=\!\frac{1}{M_h v}\left[\left(M_h^2\!+\!2 M_H^2\!-\!2 M_a^2\!-\!2 \lambda_3 v^2\right)\sin^2 \theta\!-\!2 \left(\lambda_{P1} \cos^2 \beta\!+\!\lambda_{P2}\sin^2 \beta\right)v^2 \cos^2 \theta \right] \, .
\eeq

There are strong theoretical constraints on the model, in particular conditions on the quartic Higgs couplings in order to have a scalar potential that is bounded from below \cite{Kanemura:2004mg} (similar to the case of a general 2HDM) as well as requirements from perturbative unitarity on the scattering amplitudes of Higgs into gauge boson processes~\cite{Goncalves:2016iyg}. These constraints have been discussed in e.g. Ref.~\cite{Arcadi:2019lka} and we include them in our analysis. In the limit  $M_A\gg M_a$ and for a maximal mixing  $\sin2\theta \approx 1$, these induce an  upper bound on $M_A$ of about 1.4 TeV which can, however, be  weakened by lowering the value of $\sin2\theta$. 

Let us finally introduce and discuss the DM candidate, which will be assumed to be a Dirac fermion (no substantial change of the results is expected in the case of a Majorana fermion) which is isosinglet under the SM gauge group. Because it is not charged under ${\rm SU(2)_L}$, the DM state has no couplings to gauge bosons and by virtue of the $Z_2$ symmetry which is introduced in order to make it stable, it couples to  Higgs bosons only in pairs. Starting from an initial coupling with the $a_0$ state, and after electroweak symmetry breaking, the DM will interact with the two pseudoscalar bosons according to the following Lagrangian
\begin{equation}
\mathcal{L}_{\rm DM}=g_\chi \left(\cos\theta a+\sin\theta A\right) \bar \chi i \gamma_5 \chi \, . 
\end{equation}
There are no couplings of the DM fermion to the CP-even Higgs bosons at tree-level  and this will have important consequences as we will see in the next section. 

%%%%%%%%%%%%%%%%%%%%%%%%%%%%%%%%%%%%%%%%%%%%%%%%%%%%%%%%%%%%%%%%%%%%%%%%%

\subsection*{3. Implications for collider and astroparticle physics}

\subsubsection*{3.1 Collider constraints and effects in flavor physics}

For the 2HDM part and as mentioned before, there are constraints from high-precision electroweak and SM-Higgs measurements \cite{deBlas:2021wap,Haller:2018nnx} which imply that the heavier states are approximately degenerate in mass, $M_H \simeq M_A \simeq M_{H^{\pm}} \equiv M$ and the alignment limit in which the lighter $h$ boson is SM-like. As we are interested in situations in which the couplings to isospin $-\frac12$ muons are enhanced, only the two 2HDMs of Type II and Type X (lepton-specific) with large values of $\tan\beta$, $\gsim 10$,  will be interesting for our analysis. In the special case of Type II couplings, there are LHC Higgs searches, in particular  in the processes $gg, b\bar b \to H/A \to \tau^+\tau^-$, which exclude the entire mass range $M \lsim 1$ TeV  for $\tan\beta \gsim 10$ 
\cite{Bailey:2020ecs}. There is also a severe lower bound from the decay $b\to s\gamma$ on the $H^\pm$ mass (and hence also on $M_A$ and $M_H$), $M_{H^\pm}> 570\,\mbox{GeV}$~\cite{Misiak:2017bgg}. These heavy states will thus not be discussed further in this case.  

In contrast, there are no severe bounds on the mass and couplings of the  pseudoscalar $a$  and it can be as light as a few GeV and, hence, could explain the $g-2$ anomaly as will be seen shortly. 
Searches for a light state with a mass $M_a < M_h$ have been performed in associated $a$  production with  $b\bar b$ and $\tau^+\tau^-$ pairs in $Z$ decays at LEP1 \cite{Djouadi:1990ms}. These constrain the  $a b\bar b$ and $a \tau^+ \tau^-$ couplings to be extremely tiny and smaller than those of the SM-like $h$ boson  since $Z \to bb h$ as well as $Z \to h \tau^+ \tau^-$ topologies with a light $h$ have been unsuccessfully searched for~\cite{Barate:2003sz}.  Also at LEP1, couplings of the $a$ state with gauge bosons through loops of new particles should be severely constrained by searches of the exotic $Z \to a \gamma$ decay \cite{Barate:2003sz}. Additional limits on the $Zha$ coupling come from searches in  $e^+ e^-  \to ha$ production at LEP2. 

Concerning the LHC, the most important probes of a light $a$ state come from searches in the $h \rightarrow aa$ and $h \rightarrow Za$ processes~\cite{Goncalves:2016iyg,Haisch:2018kqx,Tunney:2017yfp} which have been extensively studied. In particular, the decay $h\to aa$ for $M_a < \frac12 M_h$ is rather constraining as, for a not too small values of the $\lambda_{haa}$ coupling in eq.~(\ref{eq:haa}), it can have a significant rate given by~\cite{Ipek:2014gua,Djouadi:2005gj}
\begin{align}
\Gamma(h \rightarrow a a)= \frac{|\lambda_{haa}|^2 M_a}{8 \pi} \sqrt{1- 4 M_a^2/M_h^2 } \, . 
\end{align}
This process has been actively searched for by the ATLAS and CMS collaborations through various topologies, $2b2\mu$, $2b2\tau$, $4b$, $jj\gamma \gamma$, $2\mu 2\tau$ and $4\tau$. Besides, one can apply the general constraint on the invisible width of the SM-like Higgs boson, BR$(h\! \to \!  \mbox{inv})\!<\!0.11$ \cite{ATLAS:2020kdi}, to account for this decay and in this work, we will consider the latter more conservative approach.

Nevertheless, one can evade these bounds by choosing an extremely suppressed value of the coupling $\lambda_{haa}$. Indeed, in the 2HD+$a$ scenario (and contrary to the case of a 2HDM in the alignment limit \cite{Abe:2015oca}) one can have  such a suppression in the alignment limit and with a still sizable mixing $\sin\theta$ by a suitable  assignment of the $\lambda_{1P}$ and $ \lambda_{2P}$ couplings that leads to a partial  cancellation of some contributions in eq.~(\ref{eq:haa}). This is the approach that we adopt in our analysis in order to circumvent this constraint. 

There are also bounds on the fermionic DM state $\chi$ from missing 
transverse momentum searches at the LHC in processes in which a pseudoscalar Higgs boson is produced in association with a $W,Z$ or an $h$ boson and decays into invisible $\chi$ pairs (the pseudoscalar can also be virtual if its mass is larger than $2m_\chi$).  These mono--X signatures are to be supplemented by pseudoscalar Higgs production with pairs of heavy bottom or top quarks \cite{ATLAS:2018vvx}. 

Let us now turn to constraints from flavor physics. Such a light particle could affect a broad variety of low energy processes, in particular if it has enhanced couplings to the isospin down-type fermions.  For instance, the emission of a light $a$ enhances the decay rates of $B$ and $K$ mesons \cite{Dolan:2014ska} in Type II models. For the region of interest in the light of the $g-2$ anomaly, the most relevant processes are the decays $\Upsilon \rightarrow a \gamma$, $B_s \rightarrow \mu^+ \mu^-$ and $B \rightarrow K \mu^+ \mu^-$. The experimental bounds on these processes have been translated into constraints for the 2HD+$a$ scenario in Ref.~\cite{Arcadi:2017wqi} and will be adopted as well for the present study.

In particular, the $B_s \to \mu^+\mu^-$ decay should receive potentially large contributions from the $a$ state if it is light and has large couplings to muons. In the case of the lepton specific configuration, comparatively strong bounds as in the Type-II case can be derived by considering the searches of light leptophilic bosons recently performed by the BaBar collaboration \cite{BaBar:2020jma}. A further strong constraint comes from violation of lepton universality in the decays of the $Z$ boson and the $\tau$ lepton. We have adapted to the present 2HD+$a$ model, all the constraints determined for the 2HDM in Refs.~\cite{Abe:2015oca,Chun:2016hzs}.

%%%%%%%%%%%%%%%%%%%%%%%%%%%%%%%%%%%%%%%%%%%%%%%%%%%%%%%%%%%%%%
\subsubsection*{3.2 Dark matter constraints and requirements}

\begin{figure}[!ht]
\vspace*{-4.9cm}
    \centering
%    \subfloat{\includegraphics[width=0.29\linewidth]{feyn1.png}}~~
%     \subfloat{\includegraphics[width=0.29\linewidth]{feyn2.png}}
     \subfloat{\includegraphics[width=0.99\linewidth]{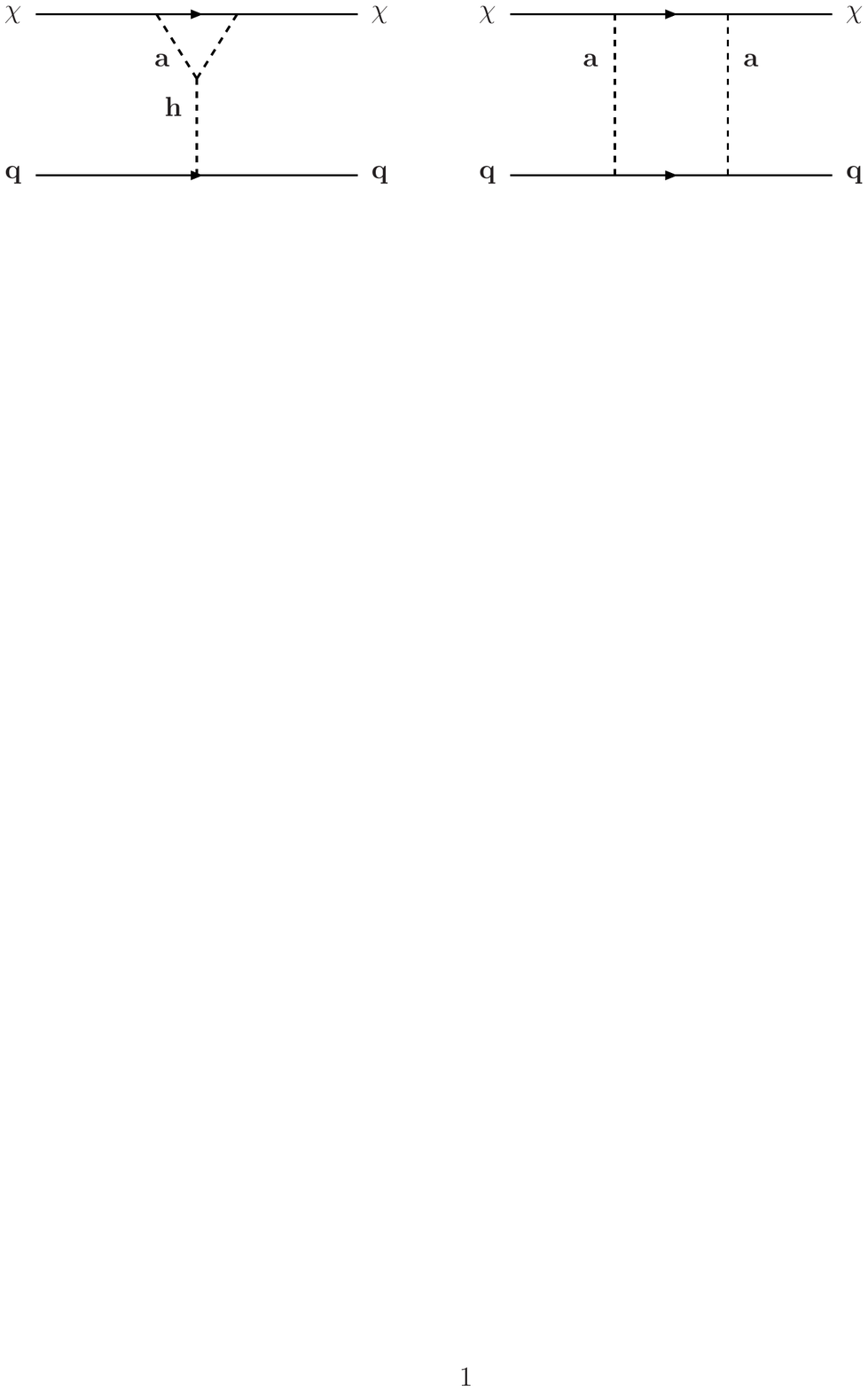}}
\vspace*{-14.5cm}
    \caption{\footnotesize{Generic Feynman diagrams responsible for the loop induced scattering of the DM state on quarks in the two Higgs doublet plus a light pseudoscalar model.}}
    \label{fig:feynloop}
\vspace*{-1mm}
\end{figure}

The 2HD+light $a$ model, as a gauge invariant embedding of a pseudoscalar portal for a SM singlet DM,  presents remarkable differences with respect to the other scenarios of fermionic DM connected to the Higgs sector. First, the absence of a coupling between the DM and the CP--even 2HDM states forbids 
spin-independent interactions for the DM at tree level. These interactions are crucial for direct detection and arise only at the one--loop level. Some Feynman diagrams which contribute to the elastic scattering of the DM with nucleons at this level are shown in Fig~\ref{fig:feynloop}. For simplicity, we have depicted only the contribution with $a$ boson exchange but all possible combinations of exchanges of the $a,A$ states should be included, although the contributions with $A$ exchange will be far smaller.  

%\textcolor{red}{
%These diagrams lead, to an effective Lagrangian relating the DM to quarks, of the form
%\begin{equation}
%\mathcal{L}=\tilde{c}_a \bar \chi \chi \bar q q,\,\,\,\,\,\tilde{c}_a=\tilde{c}_{a,\rm triangle}+\tilde{c}_{a,\rm box} \, , 
%\label{eq:eff-axx}
%\end{equation}
%and, hence,  to a spin--independent elastic cross section of the DM on protons, $\sigma_{\chi p}^{\rm SI}=\mu_{\chi p}^2/{\pi} \times |\tilde{c}_a|^2$ with $\mu_{\chi p}$ being the DM-proton reduced mass.
%}

 To compute the scattering cross section of the DM over protons, which is needed to be compared with the experimental bounds, we relied on the most recent computations performed in Refs.~\cite{Abe:2018emu,Ertas:2019dew} (see also Ref.~\cite{Arcadi:2017wqi,Bell:2018zra} for earlier estimates). The elastic DM cross-section determined in this way, has been compared with the most stringent experimental constraints as given by the XENON1T experiment \cite{XENON:2018voc}.

For what concerns the DM cosmological relic density, we will assume the conventional freeze-out paradigm in which the experimentally favored value $\Omega_{\chi}h^2 \approx 0.12$ \cite{Planck:2018vyg} is achieved via the appropriate annihilation of the DM states into SM particles. Throughout the present work, we will assume a relatively light DM particle, with a mass $m_{\chi} < m_t < M$. Under such a hypothesis, the DM relic density will be mostly accounted for by annihilation into $\tau^+  \tau^-$ and $\bar b b$ final states for Type II 2HDMs and only $\tau^+  \tau^-$ in the lepton-specific Type X model. 

The channels with $ha$, $Za$ and $aa$ final states should also be included in the annihilation subprocesses when they are kinematically accessible. Approximate expressions of the rates of these annihilation channels can be found, for example, in Ref.~\cite{Arcadi:2019lka}.

Given the fact that the DM annihilation rate into SM fermion final states is $s$-wave dominated, the model is also sensitive to constraints from DM indirect detection. We have thus included in our study the limits imposed by searches of $\gamma$-ray continuous lines as determined by the FERMI-LAT collaboration \cite{Fermi-LAT:2015att,Fermi-LAT:2015kyq} and translated them into upper limits on the annihilation cross-sections into $\tau^+ \tau^-$ and $\bar b b$ final states. 

\subsubsection*{3.3 Contributions to the g--2}

\begin{figure}[!h] 
\vspace*{-2.4cm}
    \centering
    \subfloat{\includegraphics[width=0.99\linewidth]{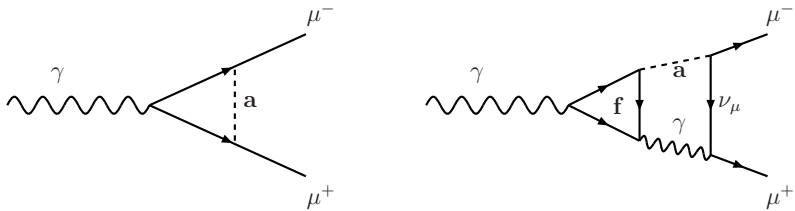}}
\vspace*{-16.7cm}
 \caption{\footnotesize{Generic Feynman diagrams responsible for the one-loop (left) and two-loop (right) contributions of a neutral pseudoscalar Higgs boson to the $(g-2)_\mu$.}}
    \label{fig:g-2}
\end{figure}

Generic neutral Higgs bosons $\Phi$ contribute to the muon $g\!-\!2$ first at the one-loop level when they are exchanged between the two muon legs\footnote{For charged Higgs bosons, there is an additional one-loop diagram in which the photon couples to the charged Higgses and a $\nu_\mu$ neutrino is exchanged between the two muons.} in the $\gamma \mu^+\mu^-$ vertex, Fig.~\ref{fig:g-2} (left). They give rise to contributions that are proportional to $m_\mu^4/ M^2_\Phi \times g^2_{\Phi \mu \mu}$ where one power of $m_\mu$ comes from the  definition, one from the kinematics and two powers come from the Yukawa couplings. When the latter are enhanced, $g_{\Phi \mu \mu} \gg 1$, the impact can be sizeable but only if the mass of the exchanged Higgs state is not too large, $M_\Phi \ll 100$ GeV.  In view of the severe bounds on the 2HDM Higgs particles from direct and indirect collider searches \cite{Haller:2018nnx}, the only state that can generate a significant contribution is the pseudoscalar $a$ boson. This occurs when  it has $i)$ a mass below the 10 GeV range, $ii)$ enhanced Yukawa couplings, meaning fermionic couplings of Type II and X with large $\tan\beta$ values, and $iii)$ a significant mixing with the $A$ boson, $\sin\theta = {\cal O}(1)$. In the limit $M_a \gg m_\mu$, one obtains at one-loop \cite{Dedes:2001nx}
\beq
\delta a_\mu^{\rm 1\!-\!loop} \approx - \frac{ \alpha}{8 \pi \sin^2\theta_W}  \frac{ m^4_\mu}{M_W^2 M_a^2} \; g_{a\mu \mu}^2 \; \bigg[ {\rm log} \bigg( \frac{M_a^2}{m_\mu^2} \bigg) - \frac{11}{6} \bigg] \, .
\eeq
Hence, in absolute value, one can indeed generate an adequate contribution to $|a_\mu|$ for $a$ masses below a few 10 GeV and $\tan\beta$ values above 20. However, because the logarithm is large and positive,  the one-loop  contribution of the $a$ state is in fact always negative and thus, cannot explain the (positive) excess observed by the Fermilab experiment, eq.~(2). 

Nevertheless, there are also possible contributions to $\delta a_\mu$ which come from  some particular Barr--Zee type diagrams \cite{Barr:1990vd}  occurring at the two-loop level \cite{Chang:2000ii,Larios:2001ma,Ilisie:2015tra} and which can be important. Indeed, as shown in the right-hand diagram of Fig.~\ref{fig:g-2}, one can generate a heavy fermion loop, $f=t,b,\tau$, that couple to the primary photon, in which a Higgs and a photon can be emitted before ending with the final muon lines\footnote{There is also the possibility of exchanging a $Z$ boson instead of the internal photon but the corresponding contribution is two orders of magnitude smaller and can be thus ignored.}. Although suppressed by two powers of the electroweak coupling, the contribution is enhanced by a factor $m_f^2/m_\mu^2$ relative to the one-loop diagram. The $a$ contribution at this level, in terms of the coupling $g_{aff}$, color number $N_c^f$ and electric charge $Q_f$ of the loop fermion $f$ with mass $m_f$, reads \cite{Chang:2000ii,Larios:2001ma,Ilisie:2015tra}
\beq
\delta a_\mu^{\rm 2\!-\!loop} = \frac{\alpha^2 }{8 \pi^2 \sin^2\theta_W} \; \frac{m_\mu^2}{M_W^2} g_{a\mu \mu}\; \sum_f g_{aff} N_c^f Q_f \; \frac{m_f^2}{M_a^2} \; F \bigg( \frac{m_f^2}{M_a^2} \bigg) \, , 
\eeq
with the function $F$ defined by 
\beq
F(r) = \int_0^1 {\rm d}x \frac{ \log (r)- \log [x(1-x)] } {r -x(1-r) } \, .
\eeq

This contribution turns out be larger than the one-loop contribution and with the correct positive sign that allows to explain the discrepancy of the measurement from the standard value. Again, this occurs for $a$ masses of a few GeV and moderately large $\tan\beta$ values which make that only closed loops of the bottom quark and the tau lepton, which also have couplings $g_{a ff} \propto \tan\beta$ in Type II and Type X scenarios, generate substantial contributions.  

\subsection*{4. Numerical analysis}

\begin{figure}[!h]
    \centering
    \includegraphics[width=0.5\linewidth]{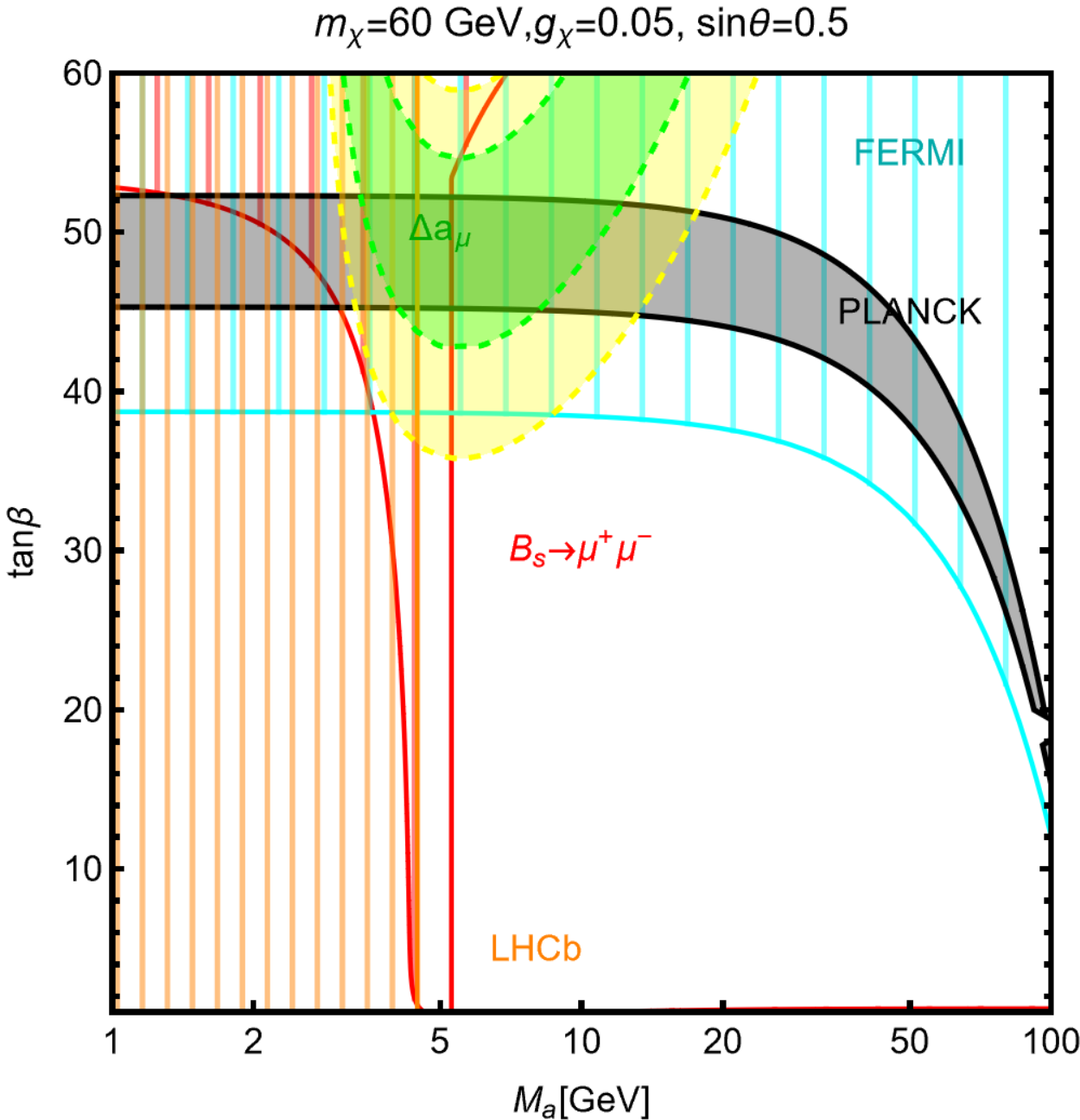}~~
    \includegraphics[width=0.5\linewidth]{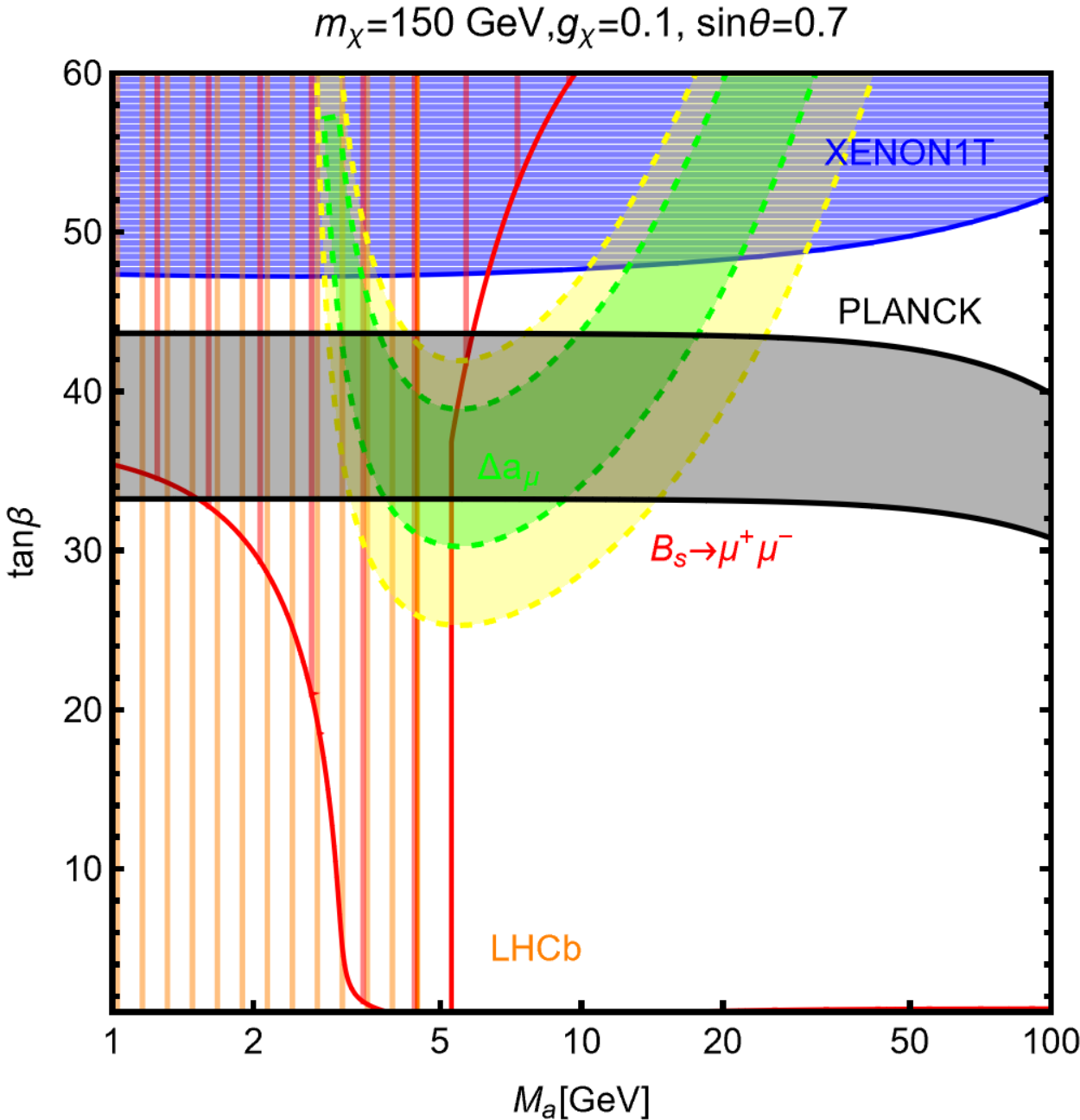}\\
    \caption{\footnotesize{The summary of constraints, in the $[M_a,\tan\beta]$ plane for the Type-II 2HDM+$a$ state for two choices of the $(m_\chi,g_\chi,\sin\theta)$ parameter set, reported on top the corresponding panels. In each plot, the colored black bands correspond to the correct DM relic density, the green (yellow) band correspond to a viable fit of the $(g-2)_\mu$ anomaly within $1\;(2)\sigma$. The red regions correspond to a rate for the $B_s \rightarrow \mu^+ \mu^-$ process exceeding the experimental determination. Finally, the blue and hatched regions correspond to the exclusion from direct DM searches by XENON1T and indirect DM searches from FERMI-LAT, while the orange region is excluded by constraints from low energy processes. For definiteness we have set the 2HDM mass scale $M$ to 1 TeV. }}
    \label{fig:p2HDMa}
\end{figure}

\begin{figure}[!h]
    \centering
    \includegraphics[width=0.5\linewidth]{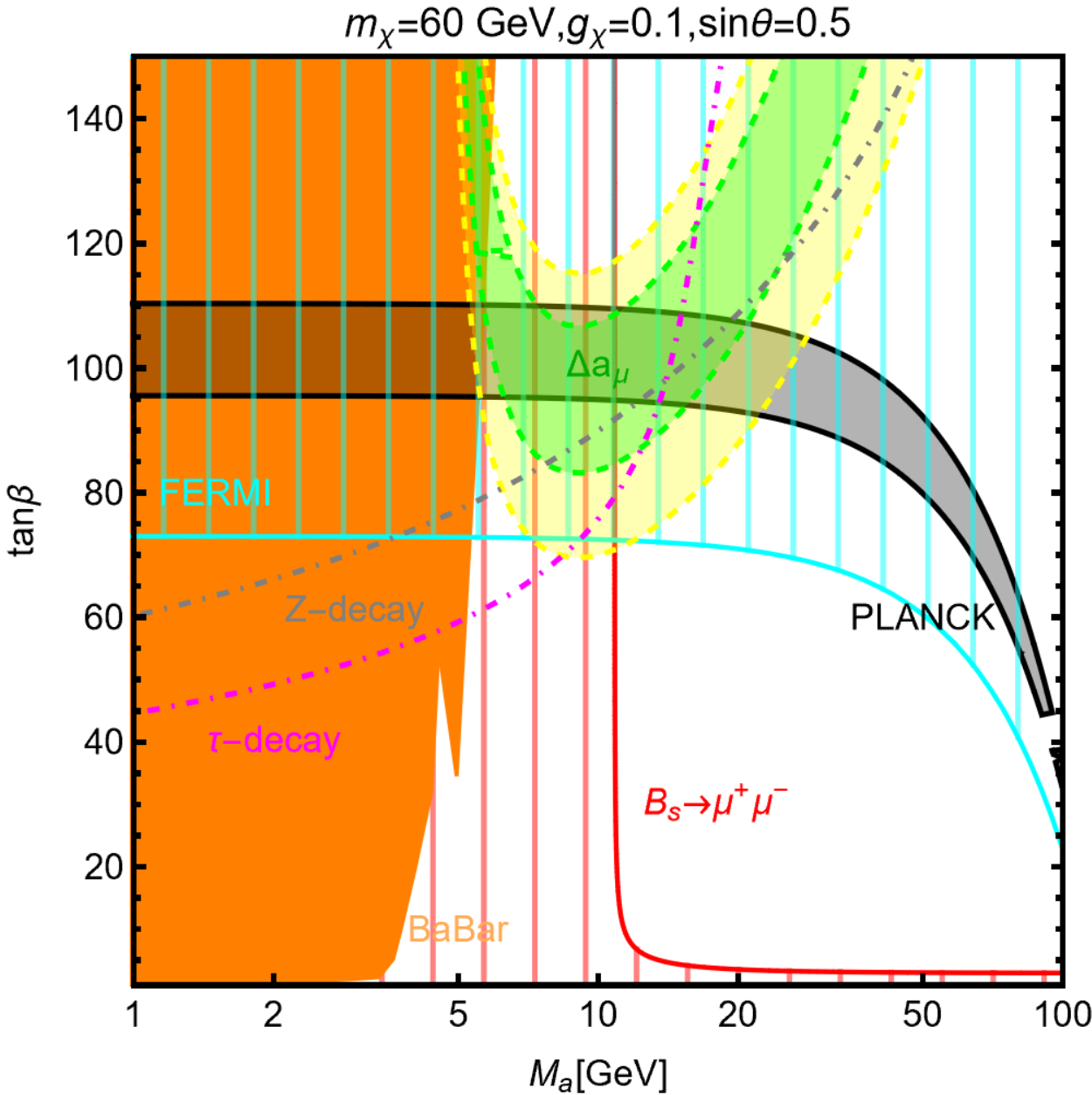}~~
    \includegraphics[width=0.5\linewidth]{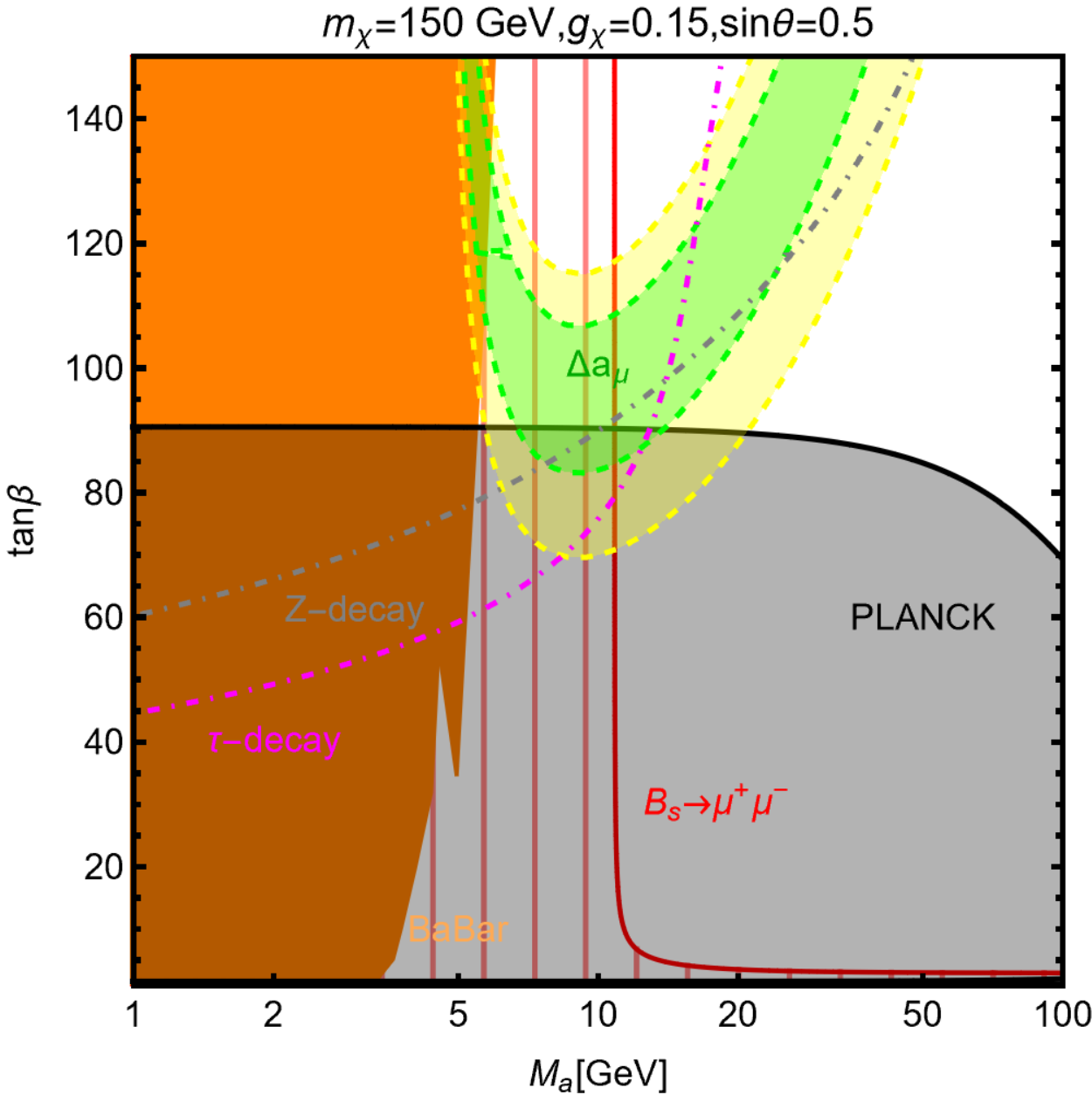}\\
    \caption{\footnotesize{The same as Fig.~\ref{fig:p2HDMa} but for the lepton-specific Type X 2HD+$a$ scenario with a 2HDM mass scale set to $M=200$ GeV. Here, we have included bounds from $Z$ and $\tau$ decays which exclude the regions above the dot-dashed gray and magenta lines.}}
    \label{fig:p2HDMaLS}
\end{figure}

We are now ready to present our numerical analysis, taking into account all the ingredients that were presented in the previous sections. Fig.~\ref{fig:p2HDMa} and Fig.~\ref{fig:p2HDMaLS} show the summary of the constraints in the $[M_a,\tan\beta]$ bidimensional plane for two benchmark scenarios in, respectively, the Type-II and Type X (lepton specific) configurations for the Higgs-fermion couplings. In each figure, we have considered two values of the DM fermion mass, namely $m_\chi=60$ GeV  and $m_\chi=150$ GeV.  In each case the values of the coupling $g_\chi$ and of $\sin\theta$ have been chosen in such a way that the correct DM relic density and a viable fit of $(g-2)_\mu$ could be achieved at the same time. In all cases we have taken $\sin\theta \gsim 0.5$, while the 2HDM $H,A$ and $H^\pm$ states are assumed to have a common mass of $M=1\,\mbox{TeV}$ in Type II and $M=200\,\mbox{GeV}$ in Type X scenarios. The DM constraints, namely from direct detection from XENON1T  and  indirect detection from FERMI, as well as flavor constraints have also been included.

As it should be evident from the figures, in Type-II and Type X scenarios, there is indeed an overlap between  the regions corresponding to the correct relic density for the DM state (in dark gray in each panel) and the regions reproducing the $(g-2)_\mu$  anomaly within $1\sigma$ (green bands) and $2 \sigma$ (yellow bands). However, these regions differ as for the considered ranges of $\tan\beta$ values. In the former case, because of the $\tan\beta$ enhancement of the Yukawa coupling of the bottom quarks, we had to impose the limit $\tan\beta \leq 60$ from the requirement of a perturbative coupling. Such a constraint is not present in the lepton-specific Type X model and, hence, higher values of $\tan\beta$ can be allowed. This feature influences strongly the allowed regions favored by the $(g-2)_\mu$ value which, indeed, tends to favor the lepton-specific scenario. In all cases, we need a sizable value of the mixing angle $\theta$, in other words require a significant doublet-like component for the pseudoscalar $a$ boson.

Besides this aspect, the Type-II and Type X models differ from the set of complementary constraints which are applied, besides the ones from the $(g-2)_\mu$ and the relic density $\Omega h^2$.  Indeed, in the former case with enhanced $g_{abb}$ couplings, one observes that the mass range  $M_a \lesssim 5\,\mbox{GeV}$ is almost entirely excluded by the bounds from $B\rightarrow K\mu \mu$ and $\Upsilon \rightarrow \gamma a$ decays (we have labeled the ensemble of the two bounds - depicted by the hatched areas in red - as ``LHCb" in the plots). Fig. \ref{fig:p2HDMa} also shows in hatched red, the regions in which the rate of $B_s \rightarrow \mu^+ \mu^-$ exceeds the experimental determination \cite{LHCb:2021awg} by more than $2\sigma$. Moving to the DM constraints, the regions of parameter space excluded by direct and indirect detection experiments are shown, respectively, as hatched areas in blue and cyan. Note that indirect detection bounds appear only for $m_\chi=60\,\mbox{GeV}$ since current experiments have not yet reached enough sensitivity to probe the freeze-out paradigm for high DM masses. 

Moving to the case of the lepton specific 2HD+$a$ case, Fig.~\ref{fig:p2HDMaLS}, we notice again that the region $M_a \lesssim 5\,\mbox{GeV}$ is excluded by searches of new light states. A comparatively stronger bound, with respect to the type-II scenario, comes from $B_s \rightarrow \mu^+ \mu^-$. This is due to the choice $M=200\,\mbox{GeV}$ which implies a sizable contribution to the rate of this processes also from other Higgs bosons than $a$. The choice of this low mass scale is needed to comply with bounds from violation of lepton universality in $Z$ decays and $\tau$ decays which are still strong. A stronger hierarchy between $M_a$ and $M$ would have completely ruled out the region corresponding to the fit of the $(g-2)_\mu$; see also Ref.~\cite{Chun:2016hzs}. 

No exclusion from direct detection experiments appears. This is due to the fact that the $1/\tan\beta$ dependence of the $a/A$ couplings to quarks causes a suppression of the contribution from the box diagram in Fig.~\ref{fig:feynloop}, while the contribution of the triangle diagram, usually dominated by the exchange of the light pseudoscalar $a$ state, is suppressed by the requirement $\lambda_{haa} \simeq 0$ to avoid a too large rate for the exotic decay of the SM-like Higgs boson into $a$ pairs. 
As can be seen, the region accounting for the  $(g-2)_\mu$ excess is tightly constrained. Nevertheless, a combined fit of $(g-2)_\mu$ and the correct relic density is still possible in the $M_a \simeq 10-20\,\mbox{GeV}$ range, for the DM mass value $m_\chi=150 \,\mbox{GeV}$. The lighter DM benchmark is, instead, again disfavored by DM indirect detection as shown in the left panel of 
Fig.~\ref{fig:p2HDMa}.

%%%%%%%%%%%%%%%%%%%%%%%%%%%%%%%%%%%%%%%%%%%%%%%%%%%%%%%%%

\subsection*{5. Conclusions}

In this note, we have studied a beyond the SM scenario in which the Higgs sector is enlarged to contain two doublets of scalar fields as well as an additional pseudoscalar Higgs boson, while the matter sector is extended by an additional electroweak isosinglet fermion which is made stable by imposing a discrete symmetry. The singlet pseudoscalar state should be rather light, would substantially mix with the heavier one of the 2HDM and one can arrange that it has strongly enhanced Yukawa couplings to isospin down-type fermions such as the muons in some scenarios dubbed Type II and Type X 2HDMs, by choosing the ratio of vacuum expectation values of the two doublet fields to be rather large,  $\tan\beta \gg 1$.   

We have shown that such a scenario first evades all the collider bounds from direct and indirect searches of new particles at LEP, the LHC and elsewhere, but also copes with all  constraints from flavor physics and provides the correct rate for the decay $B_s \rightarrow \mu^+\mu^-$. The scenario can also fulfill all the cosmological and astrophysical requirements on the additional DM fermion, namely leading to the correct cosmological DM abundance and evading the limits from direct and indirect DM detection. Finally, we have shown that for masses and couplings that are allowed by the previous constraints and requirements, one can arrange that the pseudoscalar $a$ state contributes to the muon $(g-2)$ and explains the 4.2$\sigma$  deviation of the value recently measured at Fermilab  from the one expected in the SM.

If this $(g-2)_\mu$ anomaly persists and is magnified by future more precise measurements, the 2HD+$a$ scenario could be one of the most interesting viable solutions to resolve the discrepancy as it would also address the DM issue which is very important in particle physics and cosmology. In this case, more dedicated searches for such additional Higgs and DM states should be made and these would benefit from the high-luminosity option of the LHC and the increase in sensitivity of various astroparticle experiments that are planed in the  near future.\bigskip   

\noindent {\bf Acknowledgements:}\\
AD is supported by the Estonian Research Council (ERC) grant MOBTT86 and by the Junta de Andalucia through the Talentia Senior program as well as by A-FQM-211-UGR18, P18-FR-4314 with ERDF.
FSQ have been supported by the S\~{a}o Paulo Research Foundation (FAPESP) through Grant No 2015/15897-1 and ICTP-SAIFR FAPESP grant 2016/01343-7. FSQ work was supported by the Serrapilheira Institute (grant number Serra-1912-31613), ANID – Millennium Program –  ICN2019-044, and CNPq grants {\rm 303817/2018-6} and  421952/2018-0.
%

%%%%%%%%%%%%%%%%%%%%%%%%%%%%%%%%%%%%%%%%%%%%%%%%%%%%%%%%%%%%%

\setlength{\parskip}{0.1cm}
\bibliographystyle{unsrt}
\bibliography{biblio}

\end{document}